\documentclass[11pt]{revtex4}
\usepackage{amssymb,epsf}
\usepackage{latexsym}
\usepackage{xcolor}
\usepackage{epsfig}
\usepackage{float} 
\usepackage{color} 
\usepackage{amsmath}
\usepackage{bm}
\usepackage{amsthm}
\usepackage{svg}
\usepackage{amsmath}
\usepackage{amssymb}
\usepackage{amssymb,epsf}
\usepackage{latexsym}
\usepackage{epsfig}
\usepackage{graphicx}
\usepackage{subcaption}

\usepackage[bookmarks=true,bookmarksnumbered=false,bookmarksopen=false,breaklinks=false,pdfborder={0 0 0},backref=false,colorlinks=true,citecolor=red,linkcolor=blue,urlcolor=blue]{hyperref}
\usepackage{ulem}

\begin{document}

\title{\textbf{    Classifying  deviation from standard quantum behavior using Kullback–Leibler divergence }}

\author{Salman Sajad Wani}
\email{sawa54992@hbku.edu.qa}
	\affiliation{Qatar Center for Quantum  Computing, College of Science and Engineering, \\Hamad Bin Khalifa University, Qatar}

\author{Xiaoping Shi}
\email{xiaoping.shi@ubc.ca}
\affiliation{Irving K. Barber School of Arts and Sciences, University of British Columbia Okanagan, Kelowna, BC V1V 1V7, Canada}

\author{Abrar Ahmed Naqash}
\email{abrarnaqash\_phy@nitsri.ac.in}
\affiliation{Department Of Physics, National Institute Of Technology Srinagar, Jammu and  Kashmir 190006, India}
\author{Mir Faizal}
\email{mirfaizalmir@googlemail.com}
\affiliation{Canadian Quantum Research Center, 204-3002, 32 Ave Vernon, BC V1T 2L7 Canada}
\affiliation{Irving K. Barber School of Arts and Sciences, University of British Columbia Okanagan, Kelowna, BC V1V 1V7, Canada} 

\author{Seemin Rubab}
\email{rubab@nitsri.net}
\affiliation{Department Of Physics, National Institute Of Technology Srinagar, Jammu and  Kashmir 190006, India}

\author{Yiting Chen }
\affiliation{Irving K. Barber School of Arts and Sciences, University of British Columbia Okanagan, Kelowna, BC V1V 1V7, Canada}

\author{S. Kannan}
\email{Kannan.S@anu.edu.au}
\affiliation{Department of Quantum Science and Technology, Research School of Physics, The Australian National University,
Canberra, ACT 2601, Australia.}

\author{Saif Al-Kuwari}
\email{smalkuwari@hbku.edu.qa}
\affiliation{Qatar Center for Quantum  Computing, College of Science and Engineering, \\Hamad Bin Khalifa University, Qatar}
\begin{abstract}
In this letter, we propose a novel statistical method to measure which system is better suited to probe small deviations from the usual quantum behavior. Such deviations are motivated by a number of theoretical and phenomenological motivations, and various systems have been proposed to test them. We propose that measuring deviations from quantum mechanics for a system would be easier if it has a higher  Kullback–Leibler divergence.  We show this explicitly for a non-local Scr\"{o}dinger equation and argue that it will hold for any modification to standard quantum behavior. Thus, the results of this letter can be used to classify a wide range of theoretical and phenomenological models. 
\end{abstract}
\maketitle

\section{Introduction} Several   proposals    motivated  by physically important  phenomena  predict a deviation from standard quantum behavior. For example, it is known that 
the usual  Copenhagen interpretation  makes physical phenomena    observer-dependent   \cite{1}, and to obtain   an observer-independent objective formalism of quantum mechanics,  objective-collapse theories (such as the Diosi-Penrose for the gravitational-induced collapse \cite{2}
or the collapse by stochasticity \cite{3})  have been proposed. Even though these alternatives can actually provide an observer-independent formalism for quantum mechanical decoherence, they also predict small deviations from standard quantum behavior.  
 The measurement problem has also been addressed using a deformation of the Heisenberg algebra, which again predicts a deviation from standard quantum behavior \cite{5}. 
Several experiments have been proposed to measure such deviations \cite{4, 4a, 4b,4c, 4d}, but as these deviations are very small, it is hard to detect them. In addition, there is an absence of any universal criteria to classify various experiments used to detect such small deviations; thus, it becomes hard to know which experiments should be performed. As it is hard to perform such ultra-precise experiments, it is crucial to have such criteria, which would provide important information about the effect of such deviations on the quantum systems.  It is expected that  modifications would depend not only on the kind of deviation but also on the system used to measure such a deviation. Thus, in this letter, we, for the first time, provide such a criterion to measure how the behavior of different systems changes by small modifications to standard quantum behavior. This can, in turn, act as a guiding tool for experiments to know which experiments have a better chance of detecting a specific form of modification of quantum mechanics. 
 
We point out that the modifications of quantum mechanics are not limited to the measurement problem. The modification of quantum behavior by  an intrinsic minimal length is motivated  by  quantum gravity \cite{6}. Such modifications deform the Heisenberg algebra. 
Even though such deformation of the Heisenberg algebra was initially motivated by Planck scale physics \cite{7}, they have become important in studying various low-energy effective field theories \cite{8}. In fact, it has been demonstrated that such deformations occur due to the derivative expansion of effective field theories \cite{89}. They can be used to analyze purely condensed matter effects like the consequences of next-to-nearest hopping in graphene and can modify the quantum transport in graphene at short distances \cite{9}. The criteria proposed in this paper can also find applications for analyzing and classifying such effects in condensed matter physics. Apart from such modifications, the modifications of quantum behavior also occur due to noncommutative modifications of spacetime \cite{10}, which occurs in string theory due to background fields,  and polymer quantization \cite{11} which occurs in loop quantum gravity. Various low-energy experiments have been proposed to study such deviations from quantum behavior \cite{13, 13a, 13b, 13c}. However, like the experiments used to address measurement theory \cite{4, 4a, 4b,4c, 4d}, no criteria exist to classify them, and hence we do not know which system is better suited to test such deviations from the usual quantum behavior. Here, again the results of this letter can be used, and hence the results of this letter have wide applications which go beyond the modifications motivated by measurement theory. 

We will apply the formalism developed to a physically important example to show how this formalism works. 
The specific modification will be based on a non-local modification of the original  Schr\"{o}dinger evolution \cite{12}.  However, the important point is that even though we have chosen a specific modification, similar results can be obtained for any modification of standard quantum behavior.  Even though this modification is   motivated by string field theory \cite{sft1, sft2},  such non-local modifications are not limited to string theory and occur in other approaches to quantum gravity, such as Causal sets \cite{x5}. It has also been observed that such a modification could produce entanglement in a system which was not  entangled \cite{y1, y2}. Our method can again be used to classify different experiments used to detect such behavior \cite{9, 10, 11, 12}. Non-locality is also important as it has been proposed that the non-locality is needed to restore unitarity and resolve the black hole information paradox \cite{nl}.  We will demonstrate that if we neglect the temporal modifications, then spatial non-locality 
produces corrections similar to the corrections obtained from the deformation of the  Heisenberg algebra    \cite{8, 89,  9}.    

To obtain a universal criterion to classify any  modification of quantum mechanics, we first note that   such modifications usually depend on a parameter $\beta$. We can obtain both the original and modified (by a parameter $\beta$) probability densities from their respective wave functions. Then we can use Kullback–Leibler divergence, which measures how different a probability distribution is from another \cite{kl1, kl2}. Even though the  Kullback–Leibler divergence is not a statistical distance, it can provide this important information about the separation between two distributions. If the Kullback–Leibler divergence between the original and modified probability densities is more for a given system, then it would be easier to detect such modifications than the modifications where the Kullback–Leibler divergence is small. So, the Kullback–Leibler divergence can be used to classify deviations from quantum mechanics, and we should experimentally use systems with larger Kullback–Leibler divergence.  As it is easier to study the modification of quantum mechanics using the momentum space wave function \cite{j1, j2}, we will use the  Kullback–Leibler divergence of the two divergences for a modified and original wave function in momentum space.

\section{Non-Local Deviation from Quantum Mechanics}
Even though our analysis will be valid for any theory, motivated by either quantum measurement problem \cite{1, 2,3} or quantum gravity \cite{9, 10, 11, 12}, which suggests some deviation from standard quantum behavior, we focus on a specific nonlocal model \cite{12} motivated by string field theory \cite{sft1, sft2}. However, our analysis holds for any such deviation and can be universally used to classify deviations from standard quantum behavior. 

To understand how this non-local behavior could be detected in high precision but low-velocity experiments, we start from a free complex massive scalar field, $\phi$,  with $m$ as its mass. We expect the dynamics of this field will be described by a nonlocal  Klein-Gordon equation \cite{12} due to string field theory \cite{sft1, sft2}.
\begin{equation}
 (\Box+\mu^2)\,\exp{[l_k^2(\Box+\mu^2)]}\, =0 .
\label{KGSFT}
\end{equation} 
where $\mu=mc/\hbar$ is the inverse of the reduced Compton wavelength of the field. This specific deformation occurs due to string field theory.  However, even though string field theory produces non-locality in field theories, the non-locality can also occur from other motivations \cite{x1, x2, x3, x4}, and need not be limited to string field theory. In fact, in some models, such as those obtained from  Causal sets \cite{x5}, the non-locality is not even a polynomial function. Thus, we start from a general functions $f$ of the Klein-Gordon operator $\Box = c^{-2}\partial^2_t -\nabla^2$, and write a general non-local theory as 
\begin{equation}
f(\Box+\mu^2)\phi=0
\end{equation}
 As we want to recover the standard large distance limit of the  theory, the function $f$ has to satisfy  
\begin{equation}
\lim_{l_k\to 0} f(\Box+\mu^2)\rightarrow \Box+\mu^2 
\end{equation}
where $l_{k}$ is the  scale at which non-local effects become important. 
We assume that $f$ is an analytic function, and expand it formally   as a power series   

\begin{equation}
f(z) = \sum_{n=1}^\infty b_n z^n
\end{equation}
Writting    $\phi(x)=e^{-i \frac{m c^2}{\hbar}t}\psi(t,x)$ and  taking the non-relativistic limit ($c\rightarrow\infty$), we obtain a  non-local Scr\''odinger equation \cite{12} 
\begin{equation}
\label{nlse}
f(\mathcal{S})\psi(t,x) = 0,
\end{equation}
where  $\mathcal{S}$ is defined as 
\begin{equation}
\mathcal{S}=i\hbar \frac{\partial}{\partial t} + \frac{\hbar^2}{2m}\nabla^2,
\end{equation}
The usual quantum mechanical local  Schr\"odinger equation can be obtained by   expanding  in power series, and neglecting all corrections proportional to $l_k$. However, if we consider these corrections, we can write a free non-local Schr\"odinger equation as
\begin{equation}
f(\mathcal{S})\psi = \mathcal{S}\psi +\sum_{n=2}^{\infty}b_{n}\left(\frac{-2m}{\hbar^{2}}\right)^{n-1}l_k^{2n-2}\mathcal{S}^{n}\psi =0
\label{scr}
\end{equation}
with   $b_{n}$ as dimensionless coefficients.  As the non-local deformation only occurs in the Kinetic part of the Schr\"odinger equation, we can use the non-local free Schr\"odinger equation to motivate a non-local Schr\"odinger equation with a potential term. Thus, we  can write the non-local Schr\"odinger equation in a potential $V$ as 
\begin{equation}
f(\mathcal{S})\psi = V\psi
\end{equation}
However, if we now only retain the first order corrections, and define a new constant $\beta = b_{2} ( {-2m}/{\hbar^{2}} ) l_k^{2}  $, then we observe that we obtain the equation 
\begin{equation}
    \mathcal{S}\psi + \frac{\beta}{m} \mathcal{S}^2\psi = V\psi
\end{equation}
As we usually deal with stationary states for most tabletop experiments  \cite{13, 13a, 13b, 13c}, we can neglect the temporal dependence and only keep the non-locality in spatial coordinates only. For such theories, with special non-locality, we can write the modified  Schr\"odinger equation with a $\beta \nabla^4$ correction. It is interesting to observe this is exactly the correction obtained from a deformation of the  Heisenberg algebra \cite{5, 6, 7, 8}. Thus, the modification of $[x^i, p_j] =i\hbar$ to  $[x^i, p_j] =i\hbar f^i_j(p, \beta)$, where $f^i_j(p, \beta)$  is a suitable tensorial function of the momentum and the parameter $\beta$ \cite{5}. 
This deformation was initially motivated by generalizing the usual uncertainty principle to a generalized uncertainty principle, to incorporate the existence of a minimal length that occurs in quantum gravity  \cite{6}. Here, we have observed that this can also occur from non-local field theories.  Even though various experiments have been proposed to analyze the deviation from the usual quantum behavior produced from such theories  \cite{13, 13a, 13b, 13c}, here we will for the first time use Kullback–Leibler divergence  \cite{kl1, kl2} to analyze how a specific system can be better suited to detect such deviations.   
 The Kullback–Leibler divergence measures how different a probability distribution is from another probability distribution and the more the Kullback–Leibler divergence between the original and modified probability densities $\psi^2$, the easier it would be to experimentally test them. The Kullback–Leibler divergence between an original probability density $|\psi|^2$ and a modified probability density $|\tilde \psi|^2$ is given by \cite{kl1, kl2}. 
\begin{equation}
D_{KL} (|\tilde \psi(x)|^2||  \psi (x)|^2) = \int _{-\infty}^{\infty}|\tilde\psi (x)|^2 \log\left(\frac{|\tilde \psi (x)|^2}{|\psi (x)|^2}\right)dx
\end{equation}
 We will perform our analysis in the momentum space, and to 
 define the square of wave function in the momentum space, usually the integration  measure is defined as $dp/1-\beta p^2$ \cite{j1, j2}. However, we will absorb the factor $1/1 - \beta p^2$ in the inner product of the deformed momentum wave functions, and use it to define the  Kullback–Leibler divergence measures in momentum space representation. 
Thus, we  will write the  Kullback–Leibler divergence in the momentum space as 
\begin{eqnarray}\label{eqq2}
D_{KL} (|\tilde \psi (p)|^2|  | \psi (p)|^2) = \int _{-\infty}^{\infty}|\tilde\psi (p)|^2 \log\left(\frac{|\tilde \psi (p)|^2}{|\psi (p)|^2}\right)dp
\end{eqnarray}  
We claim that this  Kullback–Leibler divergence depends on the system used to test a specific modification of quantum mechanics. We test our claim by using two simple systems of quantum mechanics modified by a deformation of the Heisenberg algebra, i.e., we will analyze a simple harmonic oscillator and a particle in a box. Even for such simple systems, with a simple modification of quantum mechanics (by spatial non-locality produced by a deformation of the Heisenberg algebra) we observe that the Kullback–Leibler divergence is different. Hence, we conclude that one of them is better suited to detect such modifications of quantum mechanics than others. 

 \section{Harmonic Oscillator}
 In the previous section, we developed a formalism to classify the deviations of quantum mechanics. In this section, we will apply this formalism to a concrete example. 
Now we analyze a quantum harmonic oscillator and obtain its Kullback–Leibler divergence. The quantum gravity corrections to the wave function for harmonic oscillator in momentum space are given by  \cite{kl3} .
\begin{equation}\label{17}
    {\Tilde{\psi}_n(p)}=2^\lambda \Gamma(\lambda) \sqrt{\frac{n!(n+\lambda)\sqrt{\beta}}{2\pi\Gamma(n+2\lambda)}} c^{1+\lambda} C_n^{\lambda}(s)
\end{equation}
where $\lambda=({1}/{2})+\sqrt{({1}/{4})+({1}/{k^4})}$, $k=\sqrt{ {\beta}{r}^{-1}}$ , $c=(1+\beta p^2)^{-1/2}$, $s=\sqrt{\beta}p(1+\beta p^2)^{-1/2} $, $r=\frac{1}{m\omega \hbar}$ and $C_n(s) $ is the  Gegnbauer polynomial. Furthermore, we assume that the ground state probability density is concentrated at the origin, which means the particle spends most of its time at the bottom of the harmonic potential well, as one would expect for a state with little energy. Thus for ground state,  $n=0$, we can write   $({\Tilde{\psi}_0(p)}=\Tilde{\psi_p})$
\begin{equation}\label{18}
     \Tilde{\psi_p}= 2^\lambda \Gamma(\lambda) \sqrt{\frac{\lambda\sqrt{\beta}}{2\pi\Gamma(2\lambda)}}\frac{1}{(1+\beta p^2)^{\frac{1+\lambda}{2}}}
\end{equation}
To find the probability density, we need to take the square modulus of the given wave function
\begin{equation}\label{19}
     |\Tilde{\psi_p}|^2=  \left(\frac{\lambda\sqrt{\beta}2^{2\lambda -1}\Gamma^2(\lambda)}{\pi\Gamma(2\lambda)}\right) \frac{1}{(1+\beta p^2)^{1+\lambda}}
\end{equation}
The probability density for the usual   Harmonic Oscillator,  can be written as \cite{kl6}
\begin{equation}\label{21}
    |\psi_p|^2=\sqrt{\frac{r}{\pi}} \exp{(-rp^2)}
\end{equation}
We note that one can arrive at the unmodified wave function Eq.\eqref{21} from the corrected wave function Eq.\eqref{19} by using the standard definition of limits 
$ 
  \displaystyle \lim _{\beta \rightarrow{0}}  \frac{1}{(1+\beta p^2)^{1+\lambda}} = e^{-r p^2}
$. 
Now we can write the Kullback–Leibler divergence between the original and modified probability densities as 
\begin{equation}\label{22}
    D_{KL} (\tilde \psi_p^2|||  \psi_p^2)= -B \int_{-\infty}^{\infty}\frac{1}{(1+\beta p^2)^{1+\lambda}}\log{\left[A(1+\beta p^2)^{1+\lambda}\exp{(-rp^2)}\right]}dp
\end{equation}
where $A$ and $B$ are given by
\begin{eqnarray}\label{AB}
    A=\frac{2^{1-2\lambda}\sqrt{r\pi}\Gamma(2\lambda)}{\sqrt{\beta}\lambda\Gamma^2(\lambda)},  
 && B=\frac{\lambda\sqrt{\beta}2^{2\lambda -1}\Gamma^2(\lambda)}{\pi\Gamma(2\lambda)}
\end{eqnarray}
Using the Gamma function–Legendre formula 
$ 
   2^{2z-1} \Gamma(z)\Gamma(z+1/2) ={\sqrt{\pi}}\Gamma(2z)
$,  we can further simplify $A$ and $B$ as
\begin{eqnarray}\label{25}
    A=\sqrt{\frac{r}{\beta}}\frac{\Gamma(\lambda+\frac{1}{2})}{\lambda\Gamma(\lambda)},  &&  B= \lambda\sqrt{\frac{\beta}{\pi}}\frac{\Gamma(\lambda)}{\Gamma(\lambda+\frac{1}{2})}
\end{eqnarray}
Finally,  we make use of Sterling  asymptotic approximation for large $z$, $ 
    \Gamma(z+1/2)=\Gamma(z) z^{1/2}
$, and   further simplify $A$ and $B$. Here, the  use of Sterling approximation is justified because       for small $\beta$, $\lambda$ takes large values 
\begin{eqnarray}
    A=\sqrt{\frac{r}{\beta\lambda }},  &&  B=\sqrt{\frac{\beta\lambda}{\pi}} 
\end{eqnarray}
 Now  binomial expanding  it in terms of $\beta$,  
$
    \lambda\approx {1}/{2}+ {r}/{\beta}+ {\beta}/{8r}
$.
Note that $\lambda$ strongly depends upon the second term only as it diverges quickly for small values $\beta$;  therefore for practical purposes we have for small $\beta$, we have 
$ 
    \lambda =  {r}/{\beta} 
$. 
Using these approximations, we can write    $A$ and $B$ as 
\begin{eqnarray}\label{30}
    A\approx 1, && 
    B\approx \sqrt{\frac{r}{\pi}} 
\end{eqnarray}

\noindent Now Eq.\eqref{22}  can be written as
\begin{equation}\label{32}
    D_{KL} (\tilde \psi_p^2|||  \psi_p^2)= -\sqrt{\frac{r}{\pi}} \int _{-\infty} ^{\infty}(1+\beta p^2)^{-\frac{r}{\beta}}\log{\left[(1+\beta p^2)^{\frac{r}{\beta}}\exp{(-rp^2)}\right]}dp
\end{equation}
Finally,  we can make use of these two Taylor's expansions, which are valid for small values of $\beta$
\begin{eqnarray}\label{33}
   (1+\beta p^2)^{-\frac{r}{\beta}}= e^{-p^2 r}+\frac{1}{2} \beta  \left(p^4 r e^{-p^2 r}\right)+O\left(\beta ^2\right)
\\
   (1+\beta p^2)^{\frac{r}{\beta}}= e^{p^2 r}-\frac{1}{2} \beta  \left(p^4 r e^{p^2 r}\right)+O\left(\beta ^2\right)
\end{eqnarray}
We can thus obtain the Kullback–Leibler divergence for a quantum harmonic oscillator for the probability density obtained from HUP (Heisenberg's Uncertainty Principle) and GUP (Generalised Uncertainty Principle) up to the strength of $\mathcal{O}(\beta^1)$ as 
\begin{equation}\label{35}
    D_{KL} (\tilde \psi_p^2|||  \psi_p^2)\approx \beta\sqrt{\frac{r}{\pi}}\int_{0 }^{\infty }   p^4 r e^{-p^2 r}  dp
\end{equation}
This integral is the standard Gaussian integral multiplied by the polynomial in the $p$ space, so we obtain an analytical solution for     $D_{KL}$ of the  harmonic oscillator as
\begin{equation}
       D_{KL} (\tilde \psi_p^2|||  \psi_p^2)\approx \frac{3}{8 r}\beta
\end{equation}
We plot this analytical solution along with the  Kullback–Leibler divergence for a particle in a box.

 \section{Particle in a Box}
 In this section, we will apply the formalism developed to a different physical system. Thus, we will be able to show that the difficulty to detect a specific modification of quantum mechanics will depend on the system chosen to detect it. This is an important observation, as it can act as a guiding principle for tabletop experiments, as it can directly give us important about the ability of such experiments to detect a specific deviation. 
 We will analyze the Kullback–Leibler divergence of a particle in a box, which can be represented by an infinite well potential. The wave function for the  $n^{th}$ excited state of this system can be written as 
$ 
    \psi(x)=\sqrt{ {2}{a}^{-1}} \sin{ {n\pi x}{a}^{-1}}
$. 
It may be noted that up to the first order in $\beta$, in the framework of spatial non-local deformation, there is no change in position space eigenstates of the quantum particle in box. However, as we have a shift of energy levels \cite{kl5}, one needs to obtain wave functions in the momentum space. The momentum‐space wave function can be obtained by the Fourier transform of the already available position‐space wave function. For the particle trapped in a one‐dimensional box, the Fourier transform is given by the following expression:
\begin{equation}\label{9}
    \psi(p)=\frac{1}{\sqrt{2\pi}}\int_0^a  \exp{(-\iota xp)}\sqrt{\frac{2}{a}} \sin{\frac{n\pi x}{a}}dx
\end{equation}
For ground state,  with  $n=1$ and setting $a=1$, we obtain
\begin{equation}
    \psi(p)=\sqrt{\pi }\left[\frac{1+\exp{(-\iota p)}}{\pi^2-p^2} \right]
\end{equation}
Now we obtain the corrected wave function in the momentum space representation. Due to the modification of the commutation algebra, we need to perform the modified Fourier Transform of our deformed wave function, which can be obtained as \cite{j1, j2}
\begin{equation}
    \Tilde{\psi}(p)=\int_0^a \sqrt\frac{1-3 \beta  p^2}{\pi}  \exp \left(-i p x \left(1-3 \beta  p^2\right)\right)\sin (\pi  x) \, dx
\end{equation}
By solving the above equation, we arrive at the deformed wave function for the infinite potential well in the momentum space representation as
\begin{equation}\label{12}
    \frac{\sqrt{\pi -3 \pi  \beta  p^2}\left(1+e^{-i p \left(1-3 \beta  p^2\right)}\right) }{\pi ^2-p^2 \left(1-3 \beta  p^2\right)^2}
\end{equation}
To find the probability density, we take the square modulus of this wave function 
\begin{equation}\label{16}
|\Tilde{\psi}(p)|^2= \frac{2 \pi  \left(1-3 \beta  p^2\right) \left(1+\cos p\left(1-3 \beta  p^2\right)\right)}{\left(\pi ^2-p^2 \left(1-3 \beta  p^2\right)^2\right)^2}
\end{equation}

To find the Kullback–Leibler divergence, we need both the modified and the unmodified probability density 
distributions.   The unmodified probability density function for infinite potential well is given by \cite{kl5}  
\begin{equation}\label{15}
    |{\psi}(p)|^2=2\pi \frac{1+\cos{p}}{(\pi^2-p^2)^2}
\end{equation}
simplifying it further, we obtain the Kullback–Leibler divergence as
\begin{equation}\label{15}
    D_{KL} (\tilde \psi_p^2|||  \psi_p^2) 
     = \int_{-\infty}^{\infty}dp \frac{2 \pi  \left(3 \beta  p^2 -1\right) \left(\cos \left(p-3 \beta  p^3\right)+1\right) \log \left(\frac{\left(\pi ^2-p^2 \left(1-3 \beta  p^2\right)^2\right)^2 (\cos (p)+1)}{\left(p^2-\pi ^2\right)^2 \left(1-3 \beta  p^2\right) \left(\cos \left(p-3 \beta  p^3\right)+1\right)}\right)}{\left(\pi ^2-p^2 \left(1-3 \beta  p^2\right)^2\right)^2}
 \end{equation}
In fact, we can  further simplify   by expanding the $\log $ term in the integral up to the first order of the $\beta$  
\begin{equation}
  D_{KL} (\tilde \psi_p^2|||  \psi_p^2) 
     p \approx \int_{-\infty}^{\infty}dp\frac{6 \pi  \beta  \left(p^5 (-\sin (p))-3 p^4-3 p^4 \cos (p)+\pi ^2 p^3 \sin (p)-\pi ^2 k^2-\pi ^2 p^2 \cos (p)\right)}{\left(\pi ^2-p^2\right)^3}
\end{equation}
As it is arduous to solve this expression analytically, we use numerical methods to plot $D_{KL}$ for infinite well potential. 
 
\begin{figure}[h]
     \centering
     \begin{subfigure}[b]{0.45\textwidth}
         \centering
         \includegraphics[width=\textwidth]{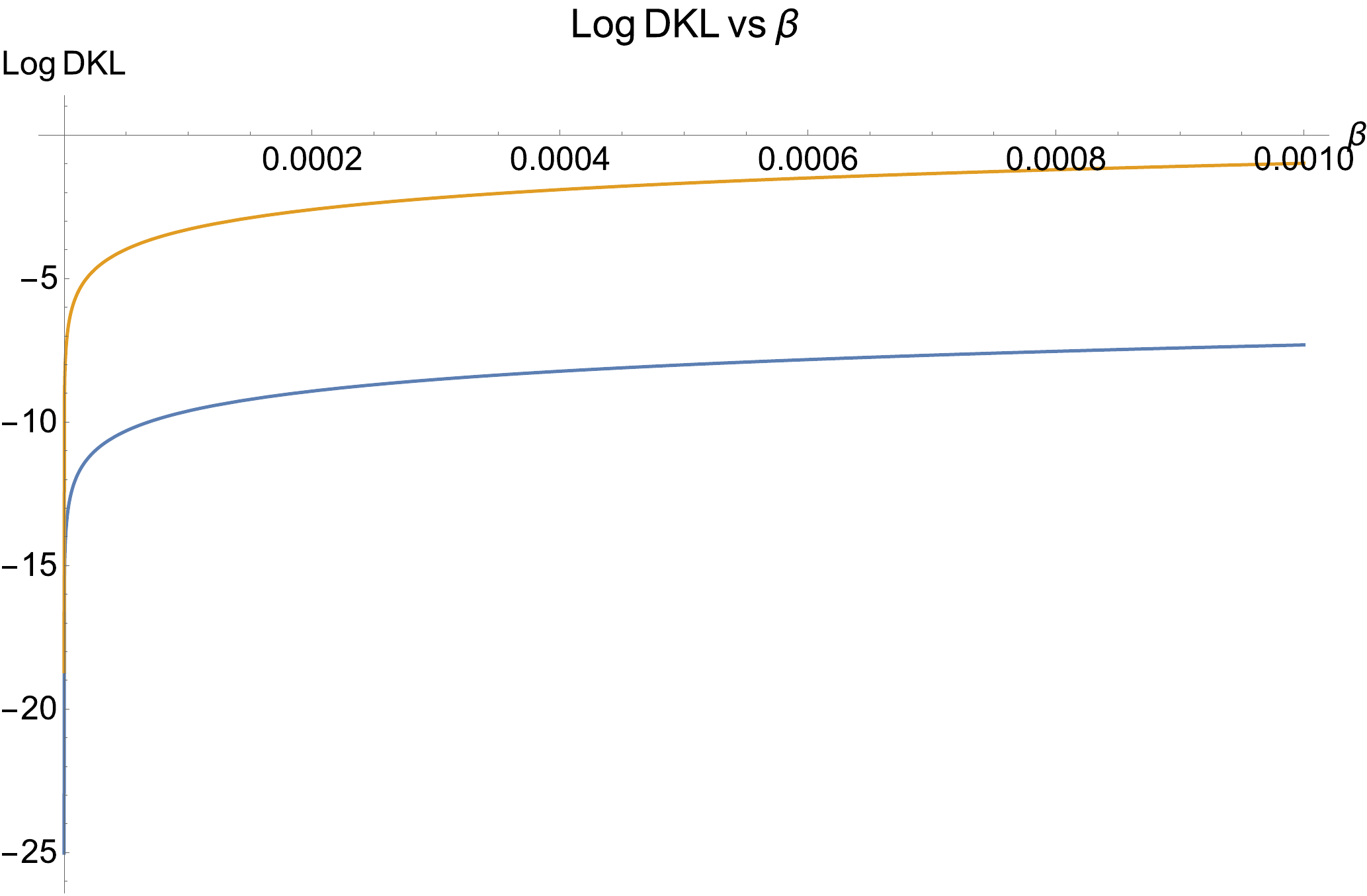}
         \caption{$\log D_{KL}$ vs $\beta$ for large values of $\beta$.}
         \label{fig:figure_labela}
     \end{subfigure}
    \hspace{0.2cm}
     \begin{subfigure}[b]{0.45\textwidth}
         \centering
         \includegraphics[width=\textwidth]{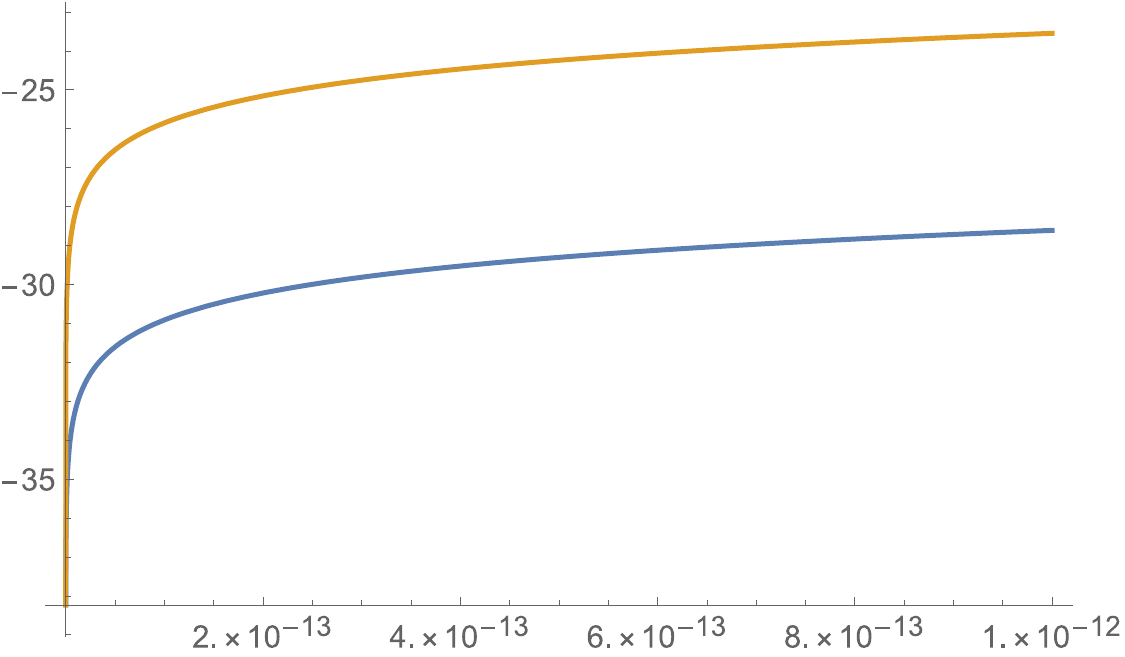}
         \caption{$\log D_{KL}$ vs $\beta$ for small values of $\beta$.}
         \label{fig:figure_labelab}
     \end{subfigure}
        \caption{log $D_{KL}$ vs $\beta$ for the quantum harmonic oscillator and  particle in the box}
        \label{fig:combined_figure}
\end{figure}

To test a modification of quantum mechanics, for a system to be able to detect small changes in the parameter $\beta$, the system should not be robust with respect to $\beta$. This can be obtained if the system has higher Kullback–Leibler divergence \cite{kl1, kl2}. Therefore, we have analyzed the Kullback–Leibler divergence  for different values of $\beta$. We have taken the length of the well equal to $a=1$  and $r=1$ for the harmonic oscillator. As the value of Kullback–Leibler divergence is small, we plot the value of $\log D_{KL}$   vs $\beta$. Figures \ref{fig:figure_labela} and \ref{fig:figure_labelab} show our results for large and small values of $\beta$, respectively.  In these figures, the orange curve describes the value of $\log D_{KL}$ for the particle in the box while the blue curve describes the same for a harmonic oscillator.
As the value of the $\log D_{KL}$  is larger for a harmonic oscillator than the particle in a box, it would be easier to experimentally test such a deviation from standard quantum behavior using a harmonic oscillator than a particle in a box. This is because the original and deformed probability densities are increasingly different from each other for a harmonic oscillator than a particle in a box. This behavior also holds for both small and large values of $\beta$.  Thus, for any experiment used to test non-locality, the deviation from standard quantum behavior will be larger for a harmonic oscillator than for a particle in a box. Hence, any experimental setup which uses the non-local modification of a harmonic oscillator will be better suited to detect non-locality than any experimental setup which uses a particle in a box. Now this behavior might be different for different modifications, and different systems, but what we have demonstrated is that the $D_{KL}$ (and hence the difficulty to experimentally detect)   can be different for various physical systems, even for the same modification. 

\section{Conclusion}
The results obtained in this  letter can have various important applications. They can classify systems that are better suited to detect a specific modification of quantum mechanics, and hence be of use to experiments in deciding which experiments would be better suited to detect a specific modification of quantum mechanics. Even though we found that for a given modification a specific system is clearly better than another to detect the deviation from ordinary quantum behavior. This may not be the case with other modifications of quantum mechanics. Hence, a thorough study of various systems and different modifications to quantum mechanics has been made to understand how different experiments can be used to probe different modifications to quantum mechanics.  Here, the important result is that Kullback–Leibler divergence can be used to classify various systems that are being used to detect a specific deviation from standard quantum behavior.

The Kullback–Leibler divergence has been used to test the robustness of the system under the change of parameter, and this is done in the sensitivity analysis of that parameter \cite{s1, s2}.  It would be interesting to analyze the robustness of $\beta$, which can be used to fix a bound on how accurately can a given system measure $\beta$. Thus, not only can the Kullback–Leibler divergence be used to classify experiments that can test a particular modification, it can provide important information on the accuracy of such tests.

\end{document}